\documentstyle[11pt,newpasp,twoside,epsf]{article}

\begin{document}

\title{Star Formation in Viscous Galaxy Disks}
\author{Adrianne Slyz,$^1$ Julien Devriendt,$^2$  Andreas Burkert,$^1$ Kevin Prendergast,$^3$ Joseph Silk$^2$}
\affil{$^1$MPIA, K\"onigstuhl 17, 69117 Heidelberg, Germany,\\ $^2$NAL, Keble Road, OX1 3RH Oxford, United Kingdom, \\ $^3$Columbia University, 10027 New York USA}


\begin{abstract}
The Lin and Pringle model (1987) of galactic disk formation postulates
that if star formation proceeds on the same timescale as the viscous
redistribution of mass and angular momentum in disk galaxies, then
the stars attain an exponential density profile. Their claim is that this
result holds generally: regardless of the disk galaxy's initial gas and
dark matter distribution and independent of the nature of the viscous
processes acting in the disk.  We present new results from a set of 2D
hydro-simulations which investigate their analytic result.
\end{abstract}

\vspace{-1.2cm}   

\section{Introduction}

A popular explanation for exponential stellar light profiles of
both high and low surface brightness disk galaxies is that stars
form {\em in situ} in an already exponential gaseous disk 
(Mo, Mao \& White 1998 and references therein).
The origin of the exponential gaseous disk is in turn
explained by the idea that all gas destined to settle into 
the disk originated in a rigidly rotating spherical protocloud 
which conserved its specific
angular momentum not only during the process of protogalactic collapse
and settling into centrifugal equilibrium but also throughout the
star formation process (Mestel 1963; Gunn 1982).  Given the 
hierarchical picture of galaxy formation which makes such 
idealized conditions unlikely, a more appealing idea is that stars 
acquire an exponential stellar profile, 
regardless of initial conditions prior to star formation, 
if star formation proceeds on a viscous timescale 
(Silk \& Norman 1981; Lin \& Pringle 1987).

We extend analytical work done by previous authors by performing 
numerical hydrodynamical simulations with 
the BGK (Bhatnagar-Gross-Krook) code (Prendergast \& Xu 1993;
Slyz \& Prendergast 1999). 
Because this code solves a model
to the collisional Boltzmann equation, it incorporates dissipation
naturally. 
Our simulations start with a gaseous disk of arbitrary initial
density profile in inviscid centrifugal equilibrium in a static dark
matter halo also of arbitrary profile. 
On a 201 by 201 evenly spaced Cartesian grid, we follow the evolution of
disks with Milky Way type masses and sizes for about 12 Gyr.
We assume an isothermal equation of state and we neglect self-gravity
and stellar feedback in this contribution.  

\vspace{-.6cm}
\begin{figure*}[!t]
\plottwo{slyza1.eps}{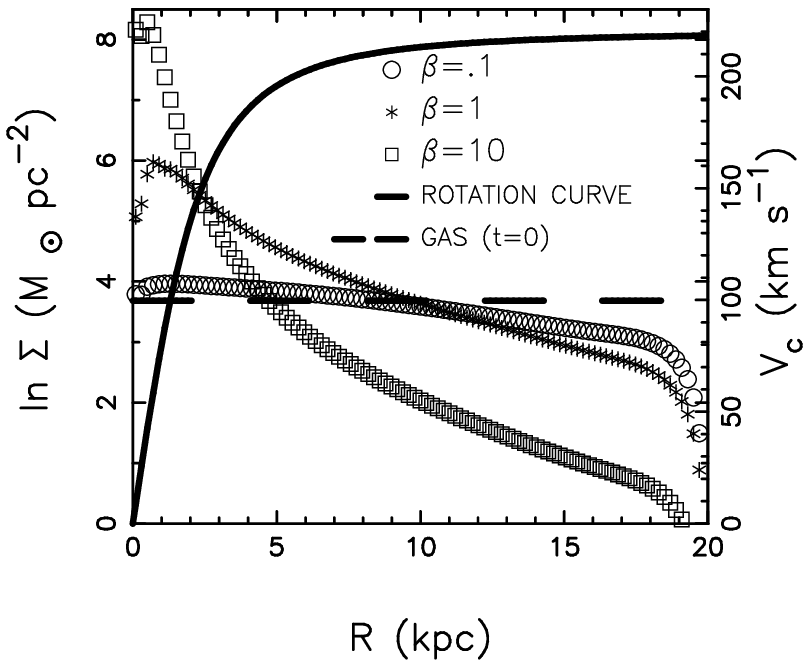}
\begin{flushleft}
\vspace{-6.85cm}
\parbox[t]{7cm}{
\caption{Comparison of stellar profiles after $\sim$ 12 Gyr for different
values of $\beta= t_\star/t_{\rm {visc}}$. 
The gravitational potential, $\Phi$, is $\frac{1}{2} {{\rm v_0}}^2 {\rm ln}({{\rm r_0}}^2 + {\rm r}^2)$ ($v_0 = 220$ km/s, $r_0=2.56$ kpc).
The initial gas profile is nearly frozen into the  stellar profile for small
$\beta$ whereas for large $\beta$ too much viscous evolution occurs
and the resulting stellar profile is more of a power law than an exponential.
\label{Bild}}}
\end{flushleft}
\end{figure*}

\section{Results}
For different initial gas and 
dark matter profiles we obtain disks with approximately
exponential stellar profiles as long as $t_\star \sim t_{\rm{visc}}$ 
(Figs. 1 and 2).    
We find that
the stellar exponential scalelength is approximately half the initial
half mass radius for the gas. Hence in so far as the disks all end up
with exponential stellar profiles, they forget their initial 
conditions but the exponential scalelength remembers the degree 
of central concentration of the initial gas distribution.  
In addition to obtaining exponential stellar profiles,
we also find gas and star fractions after $\sim$ 12 Gyr that
are consistent with observations without having to invoke an
efficiency parameter for star formation.  
The exact link between star formation and viscosity remains a question.
However our results suggest that the viscous timescale is indeed the natural 
timescale for star formation.
\begin{figure*}[!t]
\plottwo{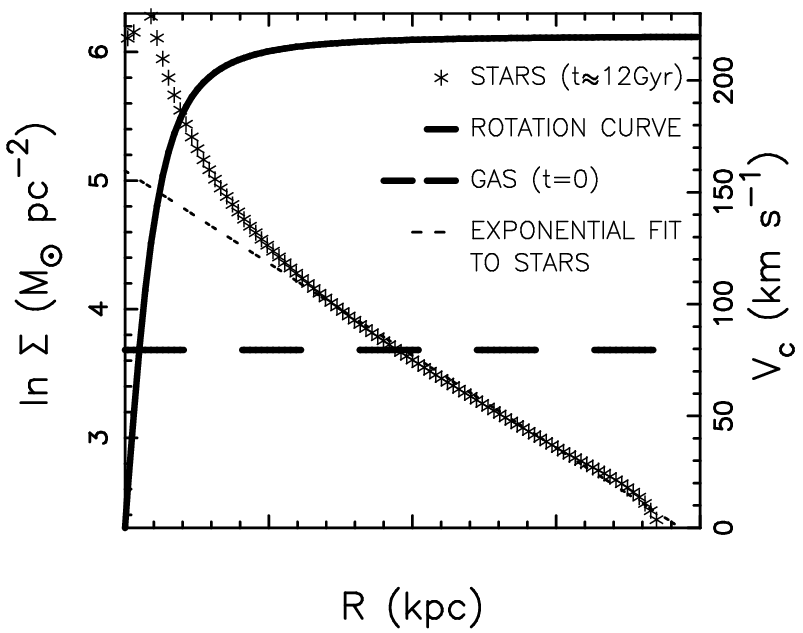}{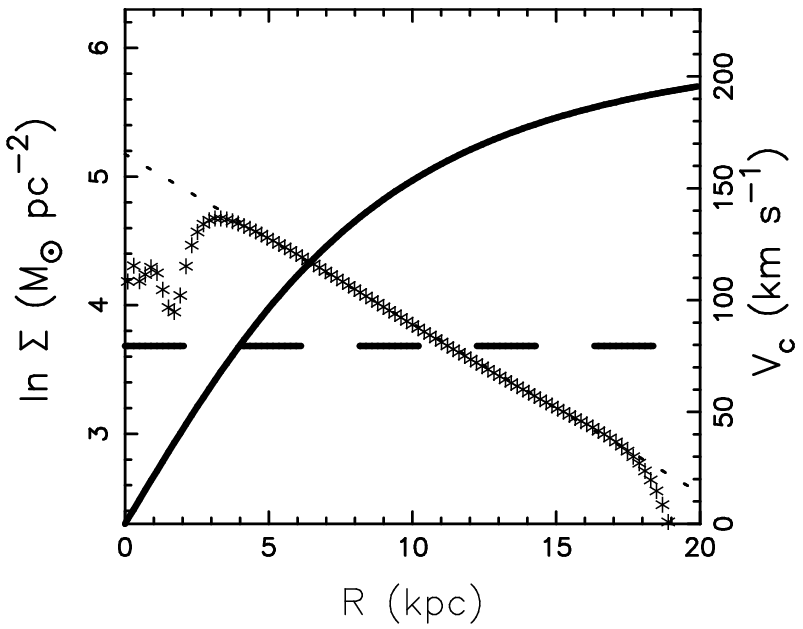}
\caption{Comparison of stellar profiles after $\sim$ 12 Gyr
for $\beta =1$ and for different rotation curves 
($r_0$ in $\Phi$ is 1.3 kpc on the left and
10.2 kpc on the right). The initial gas profile is uniform with 
$\Sigma$ = 39.8 ${\rm M}_\odot / {\rm pc^2}$. }
\end{figure*}

{\scriptsize  
}

\end{document}